\documentstyle[12pt]{article}
\let\ssection=\section
\renewcommand{\section}{\setcounter{equation}{0}\ssection}

\newcounter{def-number}[section]
\newcounter{defn}
\newenvironment{defn}{\medskip\noindent\stepcounter{def-number}
\nopagebreak[4]{\em Definition \arabic{section}.\arabic{def-number}}
\begin{enumerate}\setlength{\topsep}{-0pt}}%
{\end{enumerate}}

\newcommand{\ba}{\begin{array}}
\newcommand{\ea}{\end{array}}
\newcommand{\be}{\begin{enumerate}}
\newcommand{\ee}{\end{enumerate}}
\newcommand{\bi}{\begin{itemize}}
\newcommand{\ei}{\end{itemize}}
\newcommand{\beq}{\begin{equation}}
\newcommand{\eeq}{\end{equation}}
\newcommand{\beqa}{\begin{eqnarray}}
\newcommand{\eeqa}{\end{eqnarray}}

\newcommand{\aand}{\mbox{ and }}
\newcommand{\oor}{\mbox{ or }}
\newcommand{\eq}[1]{(\ref{#1})}

\newcommand{\ie}{{\em i.e.,\ }}

\newcommand\lemma{\smallskip\noindent{\em Lemma\ \ }\nopagebreak}
\newcommand\proof{\noindent{\em Proof\ \ }\nopagebreak} 
\newcommand\eproof{\hfill{\bf QED}\medskip}

\renewcommand{\a}{\alpha}                   
\renewcommand{\b}{\beta}                    
\newcommand{\g}{\gamma}
\newcommand{\gp}[2]{\langle#1,#2\rangle}
\newcommand{\m}{\mu}
\newcommand{\s}{\sigma}

\newcommand{\map}{\rightarrow}
\newcommand{\up}[1]{\uparrow\!\!(#1)}
 
\newcommand\mathC{\mkern1mu\raise2.2pt\hbox{$\scriptscriptstyle|$}
                {\mkern-7mu\rm C}}
\newcommand{\mathR}{{\rm I\! R}}                

\newcommand{\B}{{\cal B}}
\newcommand{\Bd}{{\cal B}^d}
\newcommand{\D}{{\cal D}}
\newcommand{\G}{\Gamma}
\renewcommand{\H}{{\cal H}}
\newcommand{\Hom}{\mbox{Hom}}
\newcommand{\Om}{\Omega}
\newcommand{\UP}{{\cal UP}}

\newcommand{\Set}{{\rm Set}}
\newcommand{\SetB}{{\Set^{\cal B}}}
\newcommand{\V}{{\cal V}}

\newcommand\mapdown[1]{\Big\downarrow
						\rlap{$\vcenter{\hbox{$\scriptstyle#1$}}$}}
\newcommand\mapright[1]{\smash{
		\mathop{\mbox{\large{$\longrightarrow$}}}\limits^{#1}}}
\newcommand\bundle[3]{\begin{array}[t]{c}
		{#1}\\ \mapdown{#2}\\ {#3}\end{array}}
\newcommand\bundlemap[2]{\begin{array}[t]{c}
		\mapright{#1}\\ \phantom{\mapdown{}}\\\mapright{#2}\\ \end{array}}

\begin{document}

\begin{titlepage}

\hspace{8truecm}Imperial/TP/95--96/55

\hspace{8truecm}gr-qc/9607069

\begin{center}
        {\large\bf Topos Theory and Consistent Histories:\\[7pt]
			The Internal Logic of the Set of all Consistent Sets}
\end{center}
\vspace{0.8 truecm}
\begin{center}
        C.J.~Isham\footnote{email: c.isham@ic.ac.uk}\\[15pt]
        The Blackett Laboratory\\
        Imperial College of Science, Technology \& Medicine\\
        South Kensington\\
        London SW7 2BZ\\
        United Kingdom\\
\end{center}

\begin{center} 26 July 1996 \end{center}

\begin{abstract} 
A major problem in the consistent-histories approach to quantum
theory is contending with the potentially large number of consistent
sets of history propositions. One possibility is to find a scheme in
which a unique set is selected in some way. However, in this paper
we consider the alternative approach in which all consistent sets
are kept, leading to a type of `many world-views' picture of the
quantum theory.  It is shown that a natural way of handling this
situation is to employ the theory of varying sets (presheafs) on the
space $\B$ of all Boolean subalgebras of the orthoalgebra $\UP$ of
history propositions. This approach automatically includes the
feature whereby probabilistic predictions are meaningful only in the
context of a consistent set of history propositions.  More
strikingly, it leads to a picture in which the `truth values', or
`semantic values' of such contextual predictions are not just
two-valued (\ie true and false) but instead lie in a larger logical
algebra---a Heyting algebra---whose structure is determined by the
space $\B$ of Boolean subalgebras of $\UP$.

	This topos-theoretic structure thereby gives a coherent
mathematical framework in which to understand the internal logic of
the many world-views picture that arises naturally in the approach
to quantum theory based on the ideas of consistent histories.
\end{abstract}
 
\end{titlepage}

\section{Introduction}
The consistent-histories approach to standard quantum theory was
pioneered by Griffiths (1984), Omn\`es (1988a, 1988b, 1988c, 1989,
1990, 1992), Gell-Mann and Hartle (1990a, 1990b) and Hartle (1991,
1995) and was motivated in part by a desire to find an
interpretation of quantum theory that is less instrumentalist than
is that of the standard `Copenhagen' view. Such a move is
particularly desirable in the context of quantum cosmology, where
any reference to an `external observer' seems singularly
inappropriate. At a more technical level, the consistent-histories
scheme provides an attractive framework in which to develop any
views on quantum gravity where the microstructure of spacetime
itself is deemed to be the subject of quantum effects. In
particular, questions about the probabilities of various
`generalised' space-times being realised in the universe are
difficult to pose and analyse in the more traditional approaches to
quantum theory.

	The central idea in the consistent-histories scheme is that (i)
under certain conditions it is possible to assign probabilities to
{\em history\/} propositions of a system rather than---as in
standard quantum theory---only to propositions concerning properties
at a fixed time; and (ii) these probabilities refer to `the way
things are' in some---as yet rather problematic---sense rather than
to the results of possible measurements made by an observer from
outside the system. A key ingredient in the theory is the
`decoherence function'---a complex-valued function $d(\a,\b)$ of
history propositions $\a,\b$ that measures the `quantum
interference' between them. A complete\footnote{A set of history
propositions is {\em complete\/} if (i) the physical realisation of
any one of them necessarily excludes all the others; and (ii) one of
them must be realised.} set of history propositions
$C:=\{\a_1,\a_2,\ldots,\a_n\}$ is said to be (strongly) $d$-{\em
consistent\/} if $d(\a_i,\a_j)=0$ for all $i\neq j$, and under these
circumstances the probability that $\a_i$ will be realised is
identified with the real number $d(\a_i,\a_i)$ (the formalism is
such that these numbers always sum to $1$ in a consistent set).

	Whether or not this scheme really answers the interpretational
questions in quantum theory has been much debated; for example
Halliwell (1995), Dowker and Kent (1995, 1996) and Kent (1996). The
central problem is the existence of many $d$-consistent sets that
are mutually incompatible in the sense that they cannot be combined
to give a single larger set.  An analogous situation arises in
standard quantum theory but there the problem is resolved by the
ubiquitous external observer deciding to measure one observable
rather than another. However, this option is not available in the
history framework and the problem must be addressed in some way that
is internal to the theory itself.

	{\em A priori\/} there are two quite different ways of
approaching this plethora of $d$-consistent sets. The first is to
try to select a unique one that is `realised' in the actual physical
world.  This might involve some natural choice within the context of
the existing framework, or it could mean looking for a new physical
process. The former is typical of those approaches to the `many
worlds'---or, in some views, `many
minds'---interpretation\footnote{For a recent review see Butterfield
(1996).} of quantum theory in which a preferred basis
in the Hilbert space of states is used to select one special branch.
An scheme of this type is proposed in Isham and Linden (1996).

	The second option is to accept the plethora of $d$-consistent
sets as a new type of many worlds or, perhaps better, `many
world-views' interpretation of quantum theory.  Understanding what
is meant by `many worlds' in existing quantum theory is inhibited by
the difficulty of finding a proper mathematical representation of
the concept, and hence of assessing it in anything other than a
rather heuristic way. However, the situation is otherwise in the
consistent-histories formalism if the different `worlds', or
`world-views', are identified with the different $d$-consistent
sets. We shall show how the mathematical structure of the collection
of all complete sets of history propositions can be exploited to
provide a novel logic with which to interpret the probabilistic
predictions of the theory in the many world-views context where all
$d$-consistent sets are handled at once.  This involves a type of
logical algebra in which probabilistic propositions are (i)
manifestly `contextual' in regard to a complete set (which need not
necessarily be $d$-consistent); and (ii) not simply binary-valued
(\ie not just true or false). However, this algebra is {\em not\/} a
`quantum logic' in the usual sense of the phrase since it {\em is\/}
distributive. On the other hand, it is not a simple classical
Boolean algebra either; rather, it is an example of an
`intuitionistic' logic.

	Of course, the idea that probabilistic assertions must be made
in the context of a $d$-{\em consistent\/} set is not new---the
theme has run through the entire development of the histories
programme and, in particular, has been re-emphasised recently by
Griffiths (1993, 1996).  However, allowing the context to be a
general complete set and, more dramatically, the use of a
`multi-valued' logic are new departures although, as I hope to show,
they follow naturally from the mathematical structure of the
consistent-histories formalism.  Indeed, the basic idea is easy
enough to state although its ramifications lead at once to concepts
drawn from the sophisticated branch of mathematics known as
`topos\footnote{A topos is a special type of category. The relevant
details are given further on in the paper.} theory'.

	To see how topos ideas arise, suppose that $C$ is a complete set
of history propositions that is {\em not\/} $d$-consistent: what
would be the status of a probabilistic prediction made in this
context? One response is ``none at all'', but a more physically
appropriate observation is that even if $C$ is not itself
$d$-consistent it might admit a coarse-graining $C'$ (\ie a set of
propositions, each of which is a sum of propositions in $C$)
that {\em is\/} $d$-consistent and in which the probabilistic
prediction does become meaningful; in other words, by agreeing to
use less precise propositions we may arrive at a situation where
probabilities {\em can\/} be assigned meaningfully.  However, we
then note that (i) any further coarse-grainings of $C'$ will also be
$d$-consistent; and (ii) there may be many such initial choices $C'$
and the same holds for further coarse-grainings of any of them.  In
the language of topos theory this says that the collection of all
$d$-consistent coarse-grainings of $C$ forms a {\em sieve\/} on $C$
with respect to the partial-ordering induced by coarse-graining. The
main idea is to assign this sieve as the `truth value'---or, perhaps
better, the `meaning' or `semantic value'---of a proposition in the
context of $C$. The crucial fact that underpins this suggestion is
that the set of all such sieves does indeed form a logical algebra,
albeit one that contains more than just the values `true' and
`false'.

	The natural occurrence of sieves in the consistent-histories
scheme is the primary motivation for claiming that topos
theory---especially the theory of sets varying over a
partially-ordered set---is the natural mathematical tool with which
to probe the internal logic of this particular approach to quantum
theory.  Fortunately, for our purposes it is not necessary to delve
too deeply\footnote{I have deliberately avoided any serious use of
category language in the main text of the paper but have added some
more technical remarks in the footnotes.} into this---rather
abstract---branch of mathematics, and to facilitate the exposition the
paper starts with a short summary of some of the relevant ideas
about varying sets. This is followed by a discussion of the crucial
poset\footnote{The word `poset' is an abbreviation for
`partially-ordered set'.} of all Boolean subalgebras of the quantum
algebra of history propositions. We then come to the heart of the
paper where the appropriate sets of semantic values for propositions
in the consistent-histories programme are investigated. Some less
central technical material is relegated to the appendices.

\section{The Topos of Varying Sets}
\subsection{Second-level propositions}\label{Subsec:second-level}
We begin with the simplest of remarks. In standard set theory,
to each subset $A$ of a set $X$ there is associated a
`characteristic map' $\chi^A:X\map\{0,1\}$ defined by
\beq
		\chi^A(x):=\left\{\ba{ll}
						1&\mbox{if $x\in A$};\\	\label{Def:chiA}
						0&\mbox{otherwise}
						\ea
				\right.
\eeq
so that
\beq
			A=(\chi^A)^{-1}\{1\}.			\label{A=chiA-1}
\eeq
Conversely, any function $\chi:X\map\{0,1\}$ defines a unique subset
$A_\chi:=\chi^{-1}\{1\}$ of $X$ whose characteristic function is
equal to $\chi$.

	Next, consider a hypothetical classical theory whose basic
ingredient is a Boolean lattice $\UP$ of propositions about the
physical universe\footnote{By `universe' I mean the physical world
with all its spatio-temporal aspects. Thus we are talking about
`generalised histories', not states of affairs at a fixed time.}. A
`pure state' $\s$ of the system will give rise to a {\em
valuation\/} on $\UP$, \ie a homomorphism $V^\s:\UP\map \Om$ from
$\UP$ to the simplest Boolean algebra $\Om:=\{0,1\}$ with `0'
interpreted as `false' and `1' as true. Thus a valuation is a
characteristic map that is also a homomorphism between Boolean
algebras.

	Now let us consider what a probabilistic version of such a
theory might look like. In theories with a realist flavour---as is,
arguably, the case with the consistent-histories programme---there
is a temptation (to which I shall succumb) to interpret probability
in the sense of `propensity' rather than in terms of
(intersubjective) states of knowledge or relative frequency of
repeated measurements. In particular, the proposition ``$\a\in\UP$
is true with probability $p$'' (to be denoted by $\gp\a p$) is to be
read as saying that the state of affairs represented by $\a$ has an
`intrinsic tendency' to occur that is measured by the number
$p\in[0,1]$. Thus the referent of $\gp{\a}p$ is to be construed as
being the universe `itself' in some way rather than, in particular,
our knowledge of the universe or the results of sequences of
measurements. A proposition of this type will be labelled
`second-level by which I mean simply that it is a proposition about
the universe that itself involves a proposition $\a\in\UP$.

	At a mathematical level, we observe that to each probability
measure $\m$ on $\UP$ (a `statistical state' of the system), and for
each $p\in[0,1]$, there is associated the subset of all $\a\in\UP$
such that $\mu(\a)=p$. In turn, this gives rise to the
characteristic map $\chi^{\m,p}:\UP\map\{0,1\}$ defined by
\beq
	\chi^{\m,p}(\a):=\left\{\ba{ll}
					  1 &\mbox{if $\mu(\a)=p$;}\label{Def:chimp}\\[3pt]
					  0 &\mbox{ otherwise}
						 \ea
					\right.
\eeq
as a particular example of the situation represented by
\eq{Def:chiA}.

	Note that the characteristic map in \eq{Def:chimp} is {\em
not\/} a valuation on $\UP$---no role is played by the Boolean
structure on $\UP$ which, in this situation, is regarded purely as a
{\em set\/}.  On the other hand, we can think of the second-level
propositions $\gp\a p$ as generating a new logical algebra with
respect to which each measure $\mu$ on $\UP$ produces a genuine
$\{0,1\}$-valued valuation $V^\m$ defined by
\beq
	V^\m\gp{\a}{p}:=\left\{\ba{ll}
					  1 &\mbox{if $\mu(\a)=p$;}\label{Def:Vm}\\[3pt]
					  0 &\mbox{ otherwise.}
						 \ea
					\right.
\eeq
Thus, for example, the conjunction operation on these new
propositions is {\em defined\/} to be such that, for all $\mu$, 
\beq
	V^\m(\gp{\a}{p}\land\gp{\b}{q})
					:=\left\{\ba{ll}
	1 &\mbox{if $\mu(\a)=p$ and $\mu(\b)=q$;}\label{Def:Vmand}\\[3pt]
					  0 &\mbox{ otherwise.}
						 \ea
					\right.	
\eeq
This leads naturally to the idea of two second-level propositions
being $\m$-{\em semantically equivalent\/} if their $V^\m$
valuations are equal, and {\em semantically equivalent\/} if they
are $\m$-semantically equivalent for all measures $\m$. For example,
for all $\mu$ and all $p\in[0,1]$ we have
\beq
		V^\m\gp{\a}p = V^\m\gp{\neg\a}{1-p}
\eeq
since $\mu(\a)+\mu(\neg\a)=1$ for all $\a\in\UP$. Hence $\gp\a p$
and $\gp{\neg\a}{1-p}$ are semantically equivalent for all
$p\in[0,1]$. A more complex example is given by the result that,
for any disjoint propositions $\a$ and $\b$ (\ie $\a\land\b=0$),
\beq
	V^\m\gp{\a\lor\b}p =
		V^\m\left(\bigvee_{q\in[0,1]}\gp{\a}{p-q}\land\gp{\b}q\right)
\eeq
which arises from the fact that $\mu(\a\lor\b)=\mu(\a)+\mu(\b)$ for
any such pair of propositions. Thus we see that, if $\a\land\b=0$,
then $\gp{\a\lor\b}p$ and
$\bigvee_{q\in[0,1]}\gp{\a}{p-q}\land\gp{\b}q$ are semantically
equivalent for all $p\in[0,1]$.

	The situation in the consistent-histories programme is similar
in many respects. Once again, there is an algebra $\UP$ of `universe
propositions' although---as part of a quantum theory---it is no
longer Boolean. There are also second-level propositions of the type
$\gp{\a}p$, although the role of a probability measure
$\m:\UP\map\mathR$ is now taken by a decoherence function
$d:\UP\times\UP\map\mathC$.  However, the really significant new
features of the consistent-histories theory are that (i) a
proposition $\gp{\a}{p}$ is physically meaningful only in the {\em
context\/} of a $d$-consistent set (or, as we shall see, any
complete set) of histories; and (ii) as we shall show, the
associated truth values, or semantic values, can be regarded as
lying in an algebra that is larger than $\{0,1\}$.

\subsection{Sets through time}
As an example\footnote{A similar example has been explored by
Dummett (1959) in the form of the proposition ``A city will never be
built on this spot''; I thank Jeremy Butterfield for this
observation.} of how contexts and generalised semantic values can
arise, consider a fixed set $X$ of people who are all alive at some
initial time, and whose bodies are preserved once they die (and who
are still referred to as `people' in that state).  Thus if
$D(t)\subseteq X$ denotes the subset of dead people at any time $t$,
then as $t$ increases $D(t)$ will clearly stay constant or increase,
\ie $t_1\leq t_2$ implies $D(t_1)\subseteq D(t_2)$. Such a
parameterised family of sets $D(t)$, $t\in\mathR$, is an example of
what has been called a ``set through time'' by those working in the
foundations of topos theory; for example Lawvere (1975), Bell
(1988), MacLane and Moerdijk (1992).  

	Now suppose that some members of our population are---in
fact---immortal. Suppose also that the members of $X$ are all
philosophers with a nostalgic leaning towards logical positivism.
Then what truth value should be assigned to the proposition ``person
$x$ is mortal'' if all truth statements are required to be
verifiable in some operational sense?  If death has already occurred
by the time the proposition is asserted then, of course, the
proposition is true (assuming that the deadness of a body is
something that can be confirmed operationally).  However, if $x$ is
alive the proposition cannot be said to be true---on the assumption
that mortality of a living being cannot be verified
operationally---but neither can it be denied, since even if $x$ {\em
is\/} numbered among the immortals there is no way of showing this.
Thus we are lead to the notion of a `stage of truth' as the context
in which a proposition acquires meaning---in our case, the time
$t$---and to the idea that the truth values of a statement at a
stage $t$ may not just lie in the set $\{0,1\}$.

	Of course, a dedicated verificationist will simply insist that
the proposition ``$x$ is mortal'' is meaningless if asserted at a
time $t_0$ when $x$ is not dead. However, topos theory provides a
more positive answer that stems from the observation that there may
be a later time $t$ at which $x$ {\em does\/} die, and then of
course $x\in D(t')$ for all times $t'\geq t$. A key idea in the
theory of sets-through-time is that the `truth value'---or, perhaps
better, the `meaning' or `{\em semantic value\/}'---at the stage
$t_0$ of the proposition ``$x$ is mortal'' is {\em defined\/} to be
the set $\chi^D_{t_0}(x)$ of all later times\footnote{The example
gets a little artificial at this point in the sense that a
transtemporal view of the history of the population $X$ is needed
for the semantic values to be appreciated: clearly a job for one of
the immortals!} $t$ at which $x$ is dead:
\beq	
	\chi^D_{t_0}(x):=\{\,t\geq t_0\mid x\in D(t)\,\}.	\label{Def:chiD}  
\eeq

	Note that if $x$ never dies, \ie if he or she is immortal, then
the right hand side of \eq{Def:chiD} is just the empty set. On the
other hand, $x$ is dead at a time $t$ if and only if
\beq
		\chi^D_t(x)=\up{t}:=[t,\infty).		\label{chiDt=up(t)}
\eeq
Equivalently, at stage $t$ we have
\beq
		D(t)=(\chi^D_t)^{-1}\{\up{t}\}.	\label{D(t)=chiD(t)-1}
\eeq
When compared with \eq{Def:chiA}, the relation \eq{D(t)=chiD(t)-1}
shows that the parameterised family of maps
$\chi^D_{t_0}:X\map\Om(t_0)$, $t_0\in\mathR$, (where $\Om(t_0)$
denotes the collection of all upper sets lying above $t_0$) is the
analogue of the single characteristic function of normal set theory.

	From a logical perspective, the crucial property of this set
$\Om(t_0)$ of all possible semantic values at stage $t_0$ is that it
possesses the structure of a Heyting algebra. Thus $\Om(t_0)$ is a
distributive lattice with the property that for any $a,b\in\Om(t_0)$
there is a unique element $(a\Rightarrow b)\in\Om(t_0)$ (with the
intuitive meaning ``if $a$ {\em then\/} $b$'') satisfying
\beq
			c\leq(a\Rightarrow b)\ \mbox{ if and only if }\ 
					c\land a\leq b.
\eeq
The negation operation in such an algebra is defined by $\neg
a:=(a\Rightarrow0)$, and satisfies the relation $a\leq\neg\neg a$
for all $a$.\footnote{This is one of the main reasons why a Heyting
algebra is chosen as the formal mathematical structure that
underlies intuitionistic logic. Thus there is a strong connection
between the theory of sets through time and the logic of
intuitionism.} Indeed, it can be shown that a Heyting algebra is
Boolean if and only if $a=\neg\neg a$ for all $a$.

\subsection{Sets varying over a partially-ordered set}
A key fact for our programme is that the ideas sketched above extend
readily to the situation where the `stages of truth' are elements of
a general partially-ordered set $\cal P$; for example, Bell (1988).
In our case, $\cal P$ will be the set of Boolean subalgebras of the
space of quantum history propositions with $W_1\leq W_2$ defined to
mean $W_2\subseteq W_1$ (so that $W_2$ is a `coarse-graining' of
$W_1$). The necessary mathematical development is most naturally
expressed in the language of category theory, although for our
purposes all that is really needed is the idea that a category
consists of a collection of things called `objects'---mathematical
entities with some precisely defined internal structure---and
`morphisms' between pairs of such objects, where a morphism is a
type of structure-preserving `map' (but not necessarily in the sense
of set theory).

	The relevant category for us is the category $\Set^{\cal P}$ of
``varying sets over $\cal P$''. Here, an object $X$ is defined to be
an assignment to each $p\in\cal P$ of a set $X(p)$, and an
assignment to each pair $p\leq q$ of a map $X_{pq}:X(p)\map X(q)$
such that (i) $X_{pp}$ is the identity map on $X(p)$, and (ii) the
following relations are satisfied\footnote{What is being exploited
here is the familiar fact that a poset $\cal P$ can be regarded as a
category in its own right in which (i) the objects are the elements
of $\cal P$; and (ii) there are no morphisms $p\map q$ unless $p\leq
q$, in which case there is just one. A `varying set over $\cal P$'
is then just a functor from the category $\cal P$ to the category
`$\Set$' of normal sets.  This is closely related to the concept of
a `presheaf' on $\cal P$.}
\beq
		X_{qr}\circ X_{pq}=X_{pr}
\eeq
whenever $p\leq q\leq r$. 

	A {\em morphism\/} $\eta:Y\map X$ between two such objects $Y,X$
in $\Set^{\cal P}$ is defined to be a family of maps
$\eta_p:Y(p)\map X(p)$, $p\in\cal P$, satisfying the compatibility
conditions\footnote{If $\cal P$ is regarded as a category then
$\eta$ is a `natural transformation' between the functors $Y$ and
$X$.}
\beq
		\eta_q\circ Y_{pq}=X_{pq}\circ\eta_p
\eeq 
as shown in the commutative diagram
\beq
	\bundle{Y(p)}{\eta_p}{X(p)}\bundlemap{Y_{pq}}{X_{pq}}	
							\bundle{Y(q)}{\eta_q}{X(q)}
\eeq
In particular, a {\em subobject\/} of a varying set\footnote{The
notation does not include a specific reference to the functions
$X_{pq}:X(p)\map X(q)$, but these are understood to be implicitly
included.}  $X=\{X(p),p\in{\cal P}\}$ is a varying set
$A=\{A(p),p\in{\cal P}\}$ with the property that $A(p)\subseteq
X(p)$ for all $p\in\cal P$, and such that $A_{pq}:A(p)\map A(q)$ is
just the restriction of $X_{pq}:X(p)\map X(q)$ to the subset
$A(p)\subseteq X(p)$. These relations are captured nicely by the
commutative diagram
\beq
\bundle{A(p)}{}{X(p)}\bundlemap{A_{pq}}{X_{pq}}	
							\bundle{A(q)}{}{X(q)}\label{cd}
\eeq
where the vertical arrows are subset inclusions.

	A simple, but important, special case is when the varying set
$X(p)$ is in fact constant \ie $X(p)=X$ for all $p\in\cal P$, and
$X_{pq}$ is the identity map from $X=X(p)$ to $X=X(q)$ for all pairs
$p\leq q$. In this situation, each set $A(p)$, $p\in\cal P$, can be
regarded as a subset of the fixed set $X$, and the condition on a
varying set $A:=\{A(p),p\in{\cal P}\}$ to be a {\em bona fide\/}
subobject of $X$ is simply that
\beq
		p\leq q\mbox{ implies } A(p)\subseteq A(q).	\label{sub-ob}
\eeq

	This special case where $X(p)$ is constant also gives rise to
the varying-set analogue of a `complement' of a subset. The obvious
family of subsets of $X$ to serve as the complement of $\{A(p),
p\in{\cal P}\}$ is $\{X-A(p),p\in{\cal P}\}$, but this does not give
a proper varying set since $p\leq q$ implies $X-A(p)\supseteq
X-A(q)$, which is the wrong behaviour. It turns out that the
appropriate definition is the genuine varying set $\neg A:=\{\neg
A(p),p\in{\cal P}\}$ where
\beq
	\neg A(p):=\{x\in X\mid \forall q\geq p, x\not\in A(q)\}.
										\label{Def:negA(p)}	
\eeq
It follows that $x\not\in\neg A(p)$ if and only if there is some
$q\geq p$ such that $x\in A(q)$, and hence
\beq
	\neg\neg A(p):=\{x\in X\mid \forall q\geq p\,\exists\, r\geq q
					\mbox{ s.t. }x\in A(r)\}.
\eeq
We see that $A(p)\subseteq\neg\neg A(p)$ whereas, in normal set
theory, the double complement of a subset is
always equal to the subset itself.  This non-standard behaviour in the
varying-set theory is a reflection of the fact that the underlying
logical structure is non-Boolean (see later). 

	As in the case of sets through time, a key role is played
by the collections $\Om(p)$, $p\in\cal P$, of all upper sets lying
above $p$.  More precisely, a {\em sieve\/}\footnote{This is the
notation employed by in Bell (1988); other authors (for example,
MacLane and Moerdijk, 1992) use the term `cosieve' for what Bell
calls a `sieve', and {\em vice versa}.} on $p$ in $\cal P$ is
defined to be any subset $S$ of $\cal P$ such that if $r\in S$ then
(i) $r\geq p$, and (ii) $r'\in S$ for all $r'\geq r$.  For each
$p\in\cal P$, the set $\Om(p)$ of all sieves on $p$ can be shown to
be a Heyting algebra\footnote{The precise algebraic relations will
be given later for the specific example of interest in the
consistent-histories theory.}, and for all pairs $p\leq q$ there is
a natural map $\Om_{pq}:\Om(p)\map\Om(q)$ defined by
\beq
		\Om_{pq}(S):=S\cap\up{q}				\label{Def:Ompq}
\eeq 
where $\up{q}:=\{r\in{\cal P}\mid r\geq q\}$ is the unit element
in the Heyting algebra $\Om(q)$ (the null element is the empty set).
It is easy to see that, with the maps $\Om_{pq}$ in \eq{Def:Ompq},
$\Om:=\{\Om(p),p\in{\cal P}\}$ is a varying set over $\cal P$
and hence an object in the category $\Set^{\cal P}$.

	A very important example of the use of $\Om$ occurs if $A$ is a
subobject of the object $X$.  There is then an associated {\em
characteristic\/} morphism $\chi^A:X\map\Om$ with, at each stage
$p\in\cal P$, the `component' $\chi^A_p:X(p)\map\Om(p)$ being
defined by
\beq
	\chi^A_p(x):=\{\,q\geq p\mid X_{pq}(x)\in A(q)\,\}\label{Def:chiAp+}
\eeq
where the fact that the right hand side of \eq{Def:chiAp+} actually
{\em is\/} a sieve on $p$ in $\cal P$ follows from the defining
properties of a subobject. Thus in each `branch' of the poset going
up from $p$, $\chi^A_p(x)$ picks out the first member $q$ (the
``time till truth'') in that branch for which $X_{pq}(x)$ lies in
the subset $A(q)$, and the commutative diagram \eq{cd} then
guarantees that $X_{pr}(x)$ will lie in $A(r)$ for all $r\geq q$. In
the special case where $X(p)=X$ for all $p$,
\eq{Def:chiAp} simplifies to ({\em c.f.} \eq{Def:chiD})
\beq
	\chi^A_p(x):=\{\,q\geq p\mid x\in A(q)\,\}\label{Def:chiAp}.
\eeq

	In what follows, the expression \eq{Def:chiAp} plays a crucial
role as the analogue in the theory of varying sets of the
characteristic map \eq{Def:chiA} $\chi^A:X\map\{0,1\}$ of normal set
theory.  Indeed, the analogue of the relation \eq{A=chiA-1} for the
situation epitomised by \eq{Def:chiAp} is ({\em c.f.}
\eq{D(t)=chiD(t)-1})
\beq
		A(p)=(\chi^A_p)^{-1}\{\,\up{p}\,\}			\label{A(p)=upp}
\eeq
at each stage $p\in\cal P$. Conversely, each morphism
$\chi:X\map\Om$ defines a subobject of $X$ (via
\eq{A(p)=upp}), and for this reason the object $\Om$ in $\Set^{\cal
P}$ is known as the {\em subobject classifier\/} in the category
$\Set^{\cal P}$; the existence of such an object is one of the
defining properties\footnote{Another defining property for a
category $\cal C$ to be a topos is that a product $A\times B$ exists
for any pair of objects $A,B$ in $\cal C$. For the full definition
see one of the standard texts (for example Bell 1988, MacLane and
Moerdijk 1992)} for a category to be a topos, which $\Set^{\cal P}$
is. As the target of characteristic maps (\ie the analogue of
$\{0,1\}$ in normal set theory), $\Omega$ can be thought of as the
`object of truth values'---an assignation that is reinforced by the
observation that $\Omega$ has an internal structure of a Heyting
algebra. For example, the conjunction $\land:\Om\times\Om\map\Om$ is
defined to be the morphism in the category $\Set^{\cal P}$ whose
components $\land_p:\Om(p)\times\Om(p)\map\Om(p)$, $p\in\cal P$, are
the conjunctions $\land_p$ in the `local' Heyting algebras $\Om(p)$;
the other logical operations are defined in a similar way.

	The main thesis of the present paper is that a situation closely
analogous to the one sketched above arises naturally in the theory
of consistent histories where the basic quantum ingredients are an
orthoalgebra $\UP$ of history propositions and a specific
decoherence function $d$. In particular, we have (i) the idea of a
`context' or a `stage'; and (ii) the property that---at each such
stage---the truth values lie in a Heyting algebra.  As emphasised
already, the key point in the formalism of consistent histories is
that we are concerned with assertions of the type $\gp{\a}{p}$
(``the probability of the history proposition $\a\in\UP$ is $p$'')
but these only have a physical meaning in the context of a
particular $d$-consistent set of propositions to which $\a$ belongs
or, to be more precise (see later), in the context of any set of
propositions that can be coarse-grained to give a $d$-consistent set
that contains $\a$.  It is technically convenient to employ the
Boolean subalgebra of $\UP$ generated by a set of history
propositions, rather than the set itself, and in this framework my
thesis is essentially that
\bi
	\item each Boolean subalgebra $W_0$ of the set $\UP$ of all history
propositions can serve as a possible `stage'; and

	\item the truth value (or semantic value) of $\gp{\a}{p}$ at a
particular stage $W_0$ is related to the collection of all
coarse-grainings $W$ of $W_0$ that contain $\a$ and are
$d$-consistent.
\ei

	As we shall see, the implementation of this idea involves a
specific application of the idea of sets varying over a poset, and
hence we do indeed obtain a Heyting algebra of possible semantic
values at each stage.  Moreover, we will show how propositions of
the type $\gp{\a}{p}$ can be associated with $\Om$-valued morphisms;
as such, they belong to the internal logic (and, indeed, formal
language) that is associated with the topos $\SetB$ where $\B$
denotes the poset of all Boolean subalgebras of $\UP$ (see below for
details). Thus the internal logic of the topos provides a framework
for understanding the logical structure of probabilistic predictions
in a consistent-histories theory in a way that automatically
includes all possible contextual references to $d$-consistent sets.
We thereby arrive at a coherent logical structure for this
particular `many world-views' picture of quantum theory.

\section{Boolean Subalgebras of Propositions}
\subsection{The general formalism of consistent histories} 
In the general approach to the consistent-history formalism
developed by Isham (1994) and Isham and Linden (1994), the central
mathematical ingredient is a pair $(\UP,\D)$ where $\UP$ is an
orthoalgebra\footnote{An orthoalgebra $\UP$ is a partially-ordered
set with greatest element $1$ and least element $0$ and for which
there is a notion of when two elements $\a,\b$ are {\em
orthogonal\/}, denoted $\a\perp\b$. If $\a$ and $\b$ are such that
$\a\perp\b$ then they can be combined to give a new element
$\a\oplus\b\in\UP$. Furthermore, $\a\leq\b$ if and only if
$\b=\a\oplus\g$ for some $\g\in\UP$. There is also a negation
operation with $\a\oplus\neg\a=1$ (for the full definition of an
orthoalgebra see Foulis, Greechie and R\"uttimann, 1992).  It should
be noted that the structure of an orthoalgebra is much weaker than
that of a lattice.  In the latter there are two connectives $\land$
and $\lor$, both of which are defined on {\em all\/} pairs of
elements, unlike the single, partial, operation $\oplus$ in an
orthoalgebra. A lattice is a special type of orthoalgebra with
$\a\oplus\b$ being defined on disjoint lattice elements $\a,\b$ (\ie
those for which $\a\leq\neg\b$) as $\a\oplus\b:=\a\lor\b$.} of
`history propositions' and $\D$ is the space of decoherence
functions defined on this algebra (for a short summary of the scheme
see Isham, 1995).

	It should be emphasised from the outset that, in practice, the
orthoalgebra formalism is much less abstract than it might appear at
first. In particular, for any given physical system it is always
appropriate to consider the possibility that $\UP$ may simply be the
algebra $P(\V)$ of projection operators on some Hilbert space $\V$;
in this case, $\a\oplus\b$ is defined if and only if $\a\b=0$, and
then $\a\oplus\b=\a+\b$.  For example, it was shown in Isham (1994)
that the history version of normal quantum theory for, say, a finite
number of time points $\{t_1,t_2,\ldots,t_n\}$ can be cast into this
form. Specifically, the history propositions are identified as
projection operators on the tensor product space
$\H_{t_1}\otimes\H_{t_2}\otimes\cdots\otimes\H_{t_n}$ where each
$\H_{t_i}$ is a copy of the Hilbert space of states $\H$ of standard
canonical quantum theory. The extension of this idea to a continuous
time variable is discussed in Isham and Linden (1995).

	Returning to the general formalism we recall that a {\em
decoherence function\/} is a map $d:\UP\times\UP\map\mathC$ that
satisfies the following conditions:
\be
	\item {\em Hermiticity:} $d(\a,\b)=d(\b,\a)^*$ for all $\a,\b$.

	\item {\em Positivity:} $d(\a,\a)\geq 0$ for all $\a$.

	\item {\em Additivity:} if $\a\perp\b$ (\ie $\a$ and $\b$ are
orthogonal) then, for all $\g$,
$d(\a\oplus\b,\g)=d(\a,\g)+d(\b,\g)$. 

	\item {\em Normalisation:} $d(1,1)=1$.
\ee
Note that the additivity condition implies that, for all $\a\in\UP$,
\beq
			d(0,\a)=0.					\label{d0a=0}
\eeq
We also note that, as shown by Isham, Linden and Schreckenberg
(1994), in the concrete case where $\UP=P(\V)$ for some Hilbert space
$\V $ every decoherence function can be written in the form
\beq
		d(\a,\b)={\rm tr}_{\V\otimes\V}(\a\otimes\b X)
\eeq
where $X$ belongs to a precisely specified class of operators on
$\V\otimes\V$.

	Following Gell-Mann and Hartle, a finite set
$C:=\{\a_1,\a_2,\ldots,\a_N\}$ of non-zero propositions is said to
be {\em complete\/} if (i) $\a_i\perp\a_j$ for all
$i,j=1,2,\ldots,N$; (ii) the elements of $C$ are `jointly
compatible', \ie they belong to some Boolean subalgebra of $\UP$;
and (iii) $\a_1\oplus\a_2\oplus\ldots\oplus\a_N=1$. In algebraic
terms, a complete set of history propositions is simply a finite
{\em partition of unity\/} in the orthoalgebra $\UP$.
   
	It should be noted that in the history version of standard
quantum theory, the decoherence function for a particular system
depends on both the initial state and the Hamiltonian. Thus, in
general, for any specific history system the decoherence function
$d$ will be one particular element of $\D$. It must be emphasised
that only $d$-{\em consistent\/} sets of history propositions are
given an immediate physical interpretation.  A complete set $C$ of
history propositions is said to be\footnote{In what follows I shall
only consider the strong case where $d(\a,\b)$ itself vanishes,
rather than just the real part of $d(\a,\b)$. The phrase `consistent
set' will always mean `strongly' consistent in this sense.} $d$-{\em
consistent\/} if $d(\a,\b)=0$ for all $\a,\b\in C$ such that
$\a\ne\b$. Under these circumstances $d(\a,\a)$ is regarded as the
{\em probability\/} that the history proposition $\a$ is true.  The
axioms above then guarantee that the usual Kolmogoroff probability
rules are satisfied on the Boolean algebra generated by $C$.

	It is worth noting that the idea of an orthoalgebra is closely
related to that of a {\em Boolean manifold\/}: an algebra that is
`covered' by a collection of maximal Boolean subalgebras with
appropriate compatibility conditions on any pair that overlap
(Hardegree and Frazer, 1982). Being Boolean, these subalgebras of
propositions carry a logical structure that is essentially
classical: a feature of the consistent-histories scheme that was
focal in the seminal work of Griffiths and Omn\`es and that has been
re-emphasised recently by Griffiths (1993, 1996).  In the approach
outlined above, these Boolean algebras are glued together from the
outset to form an orthoalgebra $\UP$ of propositions from which the
physically interpretable subsets are selected by the consistency
conditions with respect to a chosen decoherence function.

	Any partition of unity $C:=\{\a_1,\a_2,\ldots,\a_N\}$ generates
a Boolean algebra whose elements are the finite\footnote{With
appropriate care these ideas an be generalised to include countable
sets and sums, but the details are not important here.}
$\oplus$-sums of elements of $C$ (hence the elements of $C$ are
atoms of this algebra).  If $C$ is $d$-consistent, and if
$\a:=\oplus_{i\in I_1}\a_i$ and $\b:=\oplus_{j\in I_2}\a_j$ are two
members of the algebra (where $I_1$ and $I_2$ are subsets of the
index set $\{1,2,\ldots,N\}$) then, by the additivity property of
the decoherence function $d$,
\beq
	d(\a,\b)=\sum_{i\in I_1\cap I_2}d(\a_i,\a_i).
\eeq
On the other hand
\beq
	\a\land\b=\bigoplus_{i\in I_1\cap I_2}\a_i
\eeq
and so
\beq
		d(\a,\b)=d(\a\land\b,\a\land\b)		\label{dab-bool}
\eeq
for all $\a,\b$.

	In a general orthoalgebra $\UP$ not every Boolean subalgebra is
necessarily generated by a partition of unity---for example, any
Boolean subalgebra that is not atomic falls in this class. Partly
for this reason it is helpful to define consistency for Boolean
algebras {\em per se\/}, rather than go via partitions of unity.
However, it is also pedagogically useful to do so as it emphasises
the essentially `classical' nature of the properties of the
propositions in a $d$-consistent set.  The set of all Boolean
subalgebras of $\UP$ is denoted $\B$ and, motivated by
\eq{dab-bool}, we propose the following definition:

\begin{defn}
\item[] {For a given history system $(\UP,d)$, where $d\in\D$, a Boolean
subalgebra $W\in\B$ of $\UP$ is $d$-{\em consistent} if,
for all $\a,\b\in W$ we have $d(\a,\b)=d(\a\land\b,\a\land\b)$.

	Note that the smallest subalgebra $\{0,1\}$ is trivially
$d$-consistent for any $d$ since, by \eq{d0a=0}, we have
$d(0,1)=0=d(0,0)=d(0\land1,0\land1)$.  The set of all $d$-consistent
Boolean subalgebras of $\UP$ will be denoted $\Bd$.  }
\end{defn}

	The relation of this definition to the earlier one of strong
consistency is contained in the following lemma.  

\lemma The condition \eq{dab-bool} on all elements $\a,\b$ in a Boolean
subalgebra $W$ is equivalent to 
\beq
		d(\a,\b)=0\mbox{ for all }\a,\b\in W\mbox{ such that }
						\a\perp\b.			\label{dab=0}
\eeq

\proof Suppose that \eq{dab-bool} is true and let $\a\perp\b$. Then
$\a\land\b=0$ (in a Boolean algebra, $\a\perp\b$ if and only if
$\a\land\b=0$) and hence $d(\a,\b)=d(0,0)=0$.

	Conversely, suppose \eq{dab=0} is true and let $\a,\b\in W$. Then
there exist jointly orthogonal elements $\a_1,\b_1,\g_1\in W$ such
that $\a=\a_1\oplus\g$ and $\b=\b_1\oplus\g$. Indeed, we can choose
$\a_1:=\a\land\neg\b$, $\b_1:=\b\land\neg\a$ and $\g:=\a\land\b$,
where the general lattice operation $\land$ is well defined in this
Boolean subalgebra. Then, by the additivity property of the
decoherence function,
\beqa
	d(\a,\b)&=& d(\a_1\oplus\g,\b_1\oplus\g)=d(\a_1,\b_1) +
				d(\a_1,\g)+d(\g,\b_1)+d(\g,\g)		\nonumber\\
			&=&d(\a\land\b,\a\land\b)
\eeqa
since \eq{dab=0} means that $\a_1\perp\b_1$ implies
$d(\a_1,\b_1)=0$ {\em etc}.\eproof

	In what follows I shall refer to a Boolean subalgebra of $\UP$
as a {\em window\/} in order to convey the idea that it affords a
potential way of `looking' at the physical world; the initial letter
`w' also serves to remind us that a window can be regarded as a
possible `world-view', or even `weltanschauung'. A $d$-consistent
window is what Griffiths (1996) calls a `framework'.

\subsection{Key features of the space $\B$ of Boolean subalgebras}
At this point it is useful to recall a number of standard properties
possessed by $\B$ that will be important in what follows.
\be
\item {$\B$ is a partially ordered set with respect to subset
inclusion, and we write $W_1\leq W_2$ if $W_1\supseteq W_2$ (we
define $W_1\leq W_2$ in this way rather than as $W_1\subseteq W_2$
in order to be consistent with our earlier discussion of sets
varying over a poset). For such a pair $W_1\leq W_2$ we say that (i)
$W_1$ is a {\em fine-graining\/} of $W_2$; and (ii) $W_2$ is a {\em
coarse-graining\/} of $W_1$ (for convenience, this terminology
includes the idea that any window $W\in\B$ is both a coarse-graining
and a fine-graining of itself). We note that 
	\be 
	\item the greatest element of $(\B,\leq)$ is $\{0,1\}$; 
	\item there is no least element (because there exist many
 	quantum propositions $\a,\b\in\UP$ that are not compatible, 
	\ie $\a$ and $\b$ cannot be members of the same Boolean algebra) 
		but by Zorn's lemma each 
		descending\footnote{To avoid confusion note
		that a chain $\cdots\leq W_2\leq W_1\leq W_0$ of windows that 
		is descending
		with respect to the $\leq$ ordering is an {\em
		ascending\/} chain $W_0\subseteq W_1\subseteq W_2\cdots$ with 
		respect to subalgebra inclusion.} chain of windows has a minimal 
		element.  
	\ee
	}

\item A join operation can be defined on the poset $(\B,\leq)$ by
\beq
		W_1\lor W_2:=W_1\cap W_2.
\eeq

\item A general pair $W_1,W_2$ of windows will not
have a meet $W_1\land W_2$ since they will not both be subsets of
any larger Boolean algebra. However, if they {\em are\/} Boolean
compatible (\ie if there exists a Boolean algebra that contains them
both) then $W_1\land W_2$ exists as the intersection of all $W\in\B$
such that $W\supseteq W_1\cup W_2$.

\item {The set $\Bd$ of $d$-consistent windows is a 
subset of $\B$ and inherits its partial-ordering structure. In
particular, if $W_2$ is $d$-consistent then so is any $W_1\geq W_2$;
thus coarse-graining preserves $d$-consistency---a crucial property
for our later discussions.

	However, it is worth noting that if $W_1$ and $W_2$ are both
$d$-consistent and if $W_1\land W_2$ exists as a Boolean subalgebra,
it will generally not be $d$-consistent. In particular, there is
generally no biggest (with respect to $\subseteq$) $d$-consistent
coarse-graining of a non $d$-consistent window.  
	}
\ee

	Let us now recall the earlier discussion of
sets varying over a poset while making the following key definitions.

\begin{defn}
\item  A {\em sieve\/} on $W_0\in\B$ in $\B$ is a
subset (possibly empty) $S$ of windows in $\B$ such that
		\be 
			\item $W\in S$ implies $W\geq W_0$ (\ie $W\subseteq
					W_0$); 

			\item $W\in S$ and $W'\geq W$ (\ie $W'\subseteq W$)
					implies $W'\in S$.  
		\ee 
Thus a sieve on $W_0$ is an {\em upper
set\/} in $(\B,\leq)$ all of whose elements are coarse-grainings of
$W_0$. 

	The set of all sieves on $W_0$ is denoted $\Om(W_0)$.  
	
\item A sieve $S$ on $W_0$ is $d$-{\em
consistent\/} if every $W\in S$ is $d$-consistent.

\end{defn}
	
\noindent
The following properties of sieves are crucial for our purposes.
\be
	\item For each $W_0$, the set $\Om(W_0)$ of all sieves on $W_0$ in
$\B$ is a partially-ordered set with $S_1\leq S_2$ being defined as
$S_1\subseteq S_2$. The greatest element $1$ is the {\em
principal\/} sieve
		\beq
				\up{W_0}:=\{\,W\in\B\mid W\geq W_0\,\}\equiv
						\{\,W\in\B\mid W\subseteq W_0\,\}
		\eeq
and the least element $0$ is the {\em empty\/} subset of windows.

	\item {The poset $\Om(W_0)$ is a distributive lattice with the
operations (i) $S_1\land S_2:=S_1\cap S_2$; and (ii) $S_1\lor
S_2:=S_1\cup S_2$. In fact $\Om(W_0)$ is a {\em Heyting algebra\/},
\ie given sieves $S_1$ and $S_2$ there is a sieve ($S_1\Rightarrow
S_2$) such that
\beq
		S\leq (S_1\Rightarrow S_2){\rm\ if\ and\ only\ if\ }
				S\land S_1\leq S_2.	
\eeq
This sieve is defined as
\beq
	(S_1\Rightarrow S_2):=
	\{\,W\subseteq W_0\mid\forall\, W'\subseteq W
			\mbox{ if $W'\in S_1$ then $W'\in S_2$}\,\}.
										\label{Def:S1-implies-S2}
\eeq
In a Heyting algebra, the negation of an element $x$ is defined by
$\neg x:=(x\Rightarrow 0)$. Thus, for a sieve $S$ on $W_0$,
\beq
	\neg S:=\{\,W\subseteq W_0\mid\forall\, W'\subseteq W,
				 W'\not\in S\,\}.						\label{Def:negS}
\eeq

	As explained earlier, a central idea in the internal logic of
varying sets is that the Heyting algebra $\Om(W_0)$ serves as the
space of {\em semantic values\/} for propositions at the {\em
stage \/} $W_0$.
		}

	\item The collection $\Om:=\{\Om(W),W\in\B\}$ is a set
varying over $\B$ under the definition, for all pairs $W_1\leq W_2$
(\ie $W_2\subseteq W_1$),
\beqa
	\Om_{W_1W_2}:\Om(W_1)&\map&\Om(W_2)			\nonumber\\
				S\quad	 &\mapsto& S\cap\up{W_2}:=
					\{\,W\subseteq W_2\mid W\in S\,\}.
\eeqa
\ee

	We can also define the sets of $d$-consistent sieves
\beq
		\Om^d(W):=\{\,S\in\Om(W)\mid S\mbox{ is $d$-consistent} \,\}
\eeq
and note that, like $\Om$, $\Omega^d:=\{\Om^d(W),W\in\B\}$ is a set
varying over $(\B,\leq)$ if, for $W_1\leq W_2$,
$\Om^d_{W_1W_2}:\Om^d(W_1)\map\Om^d(W_2)$ is defined as
$\Om^d_{W_1W_2}(S):= S\cap\up{W_2}$.

	Finally, although no explicit use of it will be made here,
we note that another simple example of a set varying over
$(\B,\leq)$ is given by $\{\D_W,W\in\B\}$ where $\D_W$ is
defined to be the set of all decoherence functions for which $W$ is
a $d$-consistent window:
\beq
			\D_W:=\{d\in\D\mid W\in \Bd\}.
\eeq
This is a genuine object in $\SetB$ since if $W_2$ is a
coarse-graining of $W_1$ (\ie $W_1\leq W_2$) then if $d$ is such
that $W_1$ is $d$-consistent (\ie $d\in\D_{W_1})$ then $W_2$ is
$d$-consistent too (\ie $d\in\D_{W_2}$), and hence $W_1\leq W_2$
implies that $\D_{W_1}\subseteq\D_{W_2}$.

\section{Semantic Values in  Consistent Histories}
\subsection{Realisable propositions}
We come now to the main task of the paper: to formulate precisely
the idea that a second-level proposition like
$\gp{\a}{p}$ (``the probability of the history proposition $\a$
being true is $p$'') has a meaning only in the context of a window,
and with a semantic value that belongs to some logical structure
associated with that window---in particular, we anticipate that a
semantic value can be identified with a sieve on the window.

	Let us start by considering what is necessary for a proposition
$\gp{\a}{p}$ to have any meaning at all in the context of a
particular window $W$ and for a given decoherence function $d$.
Perhaps the simplest position to adopt here is that in order to be
able to `realise' $\a$ in the context of $W$, the history
proposition $\a$ must belong to the Boolean algebra $W$, and $W$
must be $d$-consistent.\footnote{Of course, it could be that
$d(\a,\a)=0$, but the statement that the state of affairs described
by the history proposition $\a$ has {\em zero\/} probability is
still a positive prediction about the universe.} This suggests the
following definition:
	
\begin{defn}
\item[] A proposition $\a\in\UP$ is $d$-{\em realisable in a
window $W$\/} if (i) $W$ is $d$-consistent; and (ii) $\a\in W$.
\end{defn}
\noindent
Then
\beq
		R^d(W):=\{\,\a\mid W\in\Bd\aand\a\in W\,\}=
						\left\{\ba{ll}
								W&\mbox{if $W\in \Bd$}\\
								\emptyset &\mbox{otherwise}
							\ea
						\right.	\label{Def:RdW}
\eeq
is the set of all\footnote{This definition has the property that
even the $0$ and $1$ history propositions are deemed not to be
$d$-realisable in a window that is not $d$-consistent. Of course,
this does affect the fact that, for all $d\in\D$, $d(0,0)=0$,
$d(0,1)=0$ and $d(1,1)=1$.} propositions that are $d$-realisable in
$W$.

	Continuing to reason in this heuristic way, we could argue next
that even if a window $W_0$ is {\em not\/} $d$-consistent, the
proposition $\gp{\a}{p}$ still has a meaning at stage $W_0$ provided
that a coarse-graining $W$ of $W_0$ exists in which $\a$ {\em is\/}
$d$-realisable. However, there may be many such coarse-grainings,
and the focal idea of the paper is that this should be reflected
by assigning an appropriate {\em semantic value\/} to
$\gp{\a}{p}$---in the present case, a natural choice would be the set of
all coarse grainings of $W_0$ in which $\a$ {\em is\/} $d$-realisable (on
the assumption that $d(\a,\a)=p$; if not, the semantic value is
deemed to be the empty set). In other words, we tentatively assign
to the second-level proposition $\gp{\a}{p}$ the semantic value at
stage $W_0$ defined as
\beq
	V^d_{W_0}\gp{\a}p:=\left\{\ba{ll}
		\{\,W\subseteq W_0\mid W\in\Bd \aand \a\in W\,\}
				&\mbox{if $d(\a,\a)=p$};\\[3pt]
			\emptyset&\mbox{otherwise}
						\ea
						\right.				\label{VpB0-1}
\eeq
which would make sense provided that the set of all such semantic
values belongs to some logical algebra.

	However, this assignment does not work in the way we have been
anticipating because the right hand side of \eq{VpB0-1} is generally
{\em not\/} a sieve on $W_0$ (because if $W$ belongs to
$V^d_{W_0}\gp{\a}p$, then any $W'\subset W$ with $\a\not\in W'$ will
{\em not\/} do so). Thus we cannot identify the set of possible
semantic values with the Heyting algebra of the space of sieves.  In
itself this does not rule out the use of \eq{VpB0-1} but it implies
that any logical structure on the set of semantic values
must be obtained in a way that is different from our anticipated use
of the topos of varying sets $\SetB$.  One possibility is sketched
in the first appendix.

\subsection{Accessible propositions}
The problem with the suggested semantic value $V^d_{W_0}\gp{\a}p$
can be seen from a somewhat different perspective by noting that
$R^d:=\{R^d(W),W\in\B\}$ does not define a proper varying set over
$\B$. This is because increasing the size of the window $W$ (\ie
fine-graining it) increases the number of propositions contained in
$W$---which suggests that $W_2\subseteq W_1$ implies
$R^d(W_2)\subseteq R^d(W_1)$---but it decreases the chances of
$d$-consistency---which suggests that $W_2\subseteq W_1$ implies
$R^d(W_2)\supseteq R^d(W_1)$. The net effect is that if $W_1\leq
W_2$ there is no obvious relation between $R^d(W_1)$ and $R^d(W_2)$
and hence no obvious candidate for the collection of maps
$R^d_{W_1W_2}:R^d(W_1)\map R^d(W_2)$ that is necessary for
$R^d=\{R^d(W),W\in\B\}$ to be a varying set.

	Note that a genuine varying set {\em can\/} be obtained by
the simple expedient of replacing the condition $\a\in W$ in
\eq{Def:RdW} by $\a\not\in W$. This gives rise to a new concept:
namely of a proposition $\a$ being `unrealisable', but this is not
what we are seeking and therefore further discussion is relegated to
an appendix where it serves as a particularly simple example of what
is meant by an object in the topos $\SetB$.

	Evidently a new idea is needed of what it means to say
that a proposition $\a$ is `realisable' in a window $W$.  On
reflection, do we really require that $\a$ actually {\em
belongs\/} to $W$? Surely it suffices if $W$ can be extended (\ie
fine-grained) to a bigger $d$-consistent window that {\em does\/}
contain $\a$?

	This new concept differs subtly from the earlier version of
`realisability' and the terminology should reflect this.  We are
thus lead to introduce a new family of second-level propositions
of the form ``$\a$ is $d$-accessible'' where $\a\in\UP$. As with
the second-level propositions $\gp{\a}{p}$, there is no reference
to windows {\em per se\/}, and I shall indicate this by referring
to propositions of this type as {\em global}.  However, the key
idea we wish to develop is that---as hinted at above in the
example of $d$-realisability---in order to interpret a global
proposition within the framework of the consistent-histories
programme it is first necessary to `localise' it by constructing
secondary versions that {\em do\/} refer to windows.  Then we can
introduce the notion of the `semantic value at a stage $W_0$' of
the global proposition, and we find that it does indeed lie in a
Heyting algebra. The aim is to use the topos-theoretic ideas
discussed earlier so that, in particular, the Heyting algebra
appropriate to a stage $W_0$ is the set $\Om(W_0)$ of sieves on
$W_0$ in $(\B,\leq)$.  

	We begin with the following definition of the `localised'
version of the new family of global propositions ``$\a$ is
$d$-accessible'', denoted $\gp{\a}{A^d}$.

\begin{defn}
\item[] A proposition $\a\in\UP$ is $d$-{\em accessible
from a window $W$\/} if there exists $W'\supseteq W$ such that
(i) $\a\in W'$; and (ii) $W'$ is $d$-consistent.\footnote{Note that this
implies that $W$ itself must be $d$-consistent---no propositions
are accessible from a non $d$-consistent window.}
\end{defn}
Then
\beq
	A^d(W):=
		\{\,\a\mid\exists\, W'\supseteq W\mbox{ s.t. }
				W'\in\Bd\aand \a\in W'\,\}	\label{Def:AdW}
\eeq
is the set of all propositions that are $d$-accessible from $W$.
Note that this can be rewritten as
\beq
	A^d(W)=\{\,\a\mid\exists\, W'\supseteq W\mbox{ s.t. }
				 \a\in R^d(W')\,\}
\eeq
where $R^d(W)$ was defined in \eq{Def:RdW}. Thus a proposition $\a$
is $d$-accessible from a window $W$ if and only if there exists some
{\em fine-graining\/} $W'$ of $W$ in which $\a$ is $d$-realisable.

	The following properties of these sets are crucial for our
purposes.
\be 
\item If $W_1\leq W_2$ then $A^d(W_1)\subseteq A^d(W_2)$,
and hence, unlike the case for $R^d$, the collection
$A^d:=\{A^d(W),W\in\B\}$ is a genuine varying set over $(\B,\leq)$
with $A^d_{W_1W_2}:A^d(W_1)\map A^d(W_2)$ defined as subset
inclusion.  In this sense, `accessibility' works while
`realisability' fails. As we shall see, this new object $A^d$ in
$\SetB$ is the crucial ingredient in my topos-based semantics for
interpreting the second-level propositions $\gp{\a}{p}$ in the
consistent-histories programme.

\item {The varying set $A^d$ is a subobject of the
constant varying set $\Delta\UP$ in $\Set^B$,
\beq
		\Delta\UP(W):=\UP \mbox{ for all $W\in\B$}.
\eeq
Hence there is a characteristic morphism
$\chi^{A^d}:\Delta\UP\map\Om$ in the topos $\SetB$ which, according to
\eq{Def:chiAp}, is defined at any stage $W_0$ by ({\em c.f.},
\eq{Def:chiD})
\beqa	
	\chi^{A^d}_{W_0}:\Delta\UP(W_0)&\map&\Om(W_0) \label{Def:chiAd}\\
		\a\quad &\mapsto&\{\,W\geq W_0\mid\a\in A^d(W)\,\}	\nonumber\\
			&\ =&\{\,W\subseteq W_0\mid\exists\, W'\supseteq W
			\mbox{ s.t. }W'\in \Bd\aand \a\in W'\,\}\nonumber
\eeqa
where, as can readily be checked, the right hand side is a {\em
bona fide\/} sieve on $W_0$.
	}
\ee

	The sieve on the right hand side of \eq{Def:chiAd}---which actually
belongs to the subset $\Om^d(W_0)$ of $\Om(W_0)$---is interpreted as
the semantic value at the stage $W_0$ of the global proposition
``$\a$ is $d$-accessible''. Note that, by the property
\eq{A(p)=upp} of a characteristic morphism, a history
proposition $\a$ is $d$-accessible from a window $W$ if and only if
\beq
	\chi^{A^d}_{W}(\a)=\up{W} \label{chiAd=upW}
\eeq 
where $\up{W}=\{W'\in\B\mid W\leq W'\}$ is the unit $1$ of
the Heyting algebra $\Om(W)$.

	The definition \eq{Def:chiAd} of $\chi^{A^d}:\Delta\UP\map\Om$
and property \eq{chiAd=upW} illustrate the essentially `fuzzy'
nature\footnote{This is not a coincidence: fuzzy set theory can be
viewed as a sub-branch of topos theory.} of subobjects in $\SetB$.
More precisely, if $\a$ is $d$-accessible from $W$ then
$\chi^{A^d}_{W}(\a)=1$; but even if $\a$ is not $d$-accessible from
a window $W_0$, the proposition $\gp{\a}{A^d}$ is still ascribed a
semantic value at stage $W_0$ that is generically not the null
element $0$ (the empty sieve) of the Heyting algebra $\Om(W_0)$:
namely, the set of all coarse-grainings $W$ of $W_0$ from which $\a$
{\em is\/} $d$-accessible. Thus the semantic value at stage $W_0$ of
the global proposition ``$\a$ is $d$-accessible'' is a measure of
the extent to which $W_0$ needs to be changed in order that $\a$
{\em does\/} become $d$-accessible from it.  Hence coarse-graining a
window is the analogue in the consistent-histories theory of
choosing a later time in the example of sets-through-time discussed
earlier (compare
\eq{chiAd=upW} with
\eq{chiDt=up(t)}).

	A key role for $\Om$ is to impart a logical structure to the
collection of all global propositions, and the first step in this
direction is to note that \eq{Def:chiAd} can be used to define what
is known in the topos literature as a {\em global element\/} of the
object $\Om$, \ie a morphism $1\map\Om$ in $\SetB$ where $1$ is the
terminal object\footnote{An object $1$ is said to be a {\em terminal
object\/} in a category if there is just one morphism from any other
object to $1$; it is easy to see that any two terminal objects are
isomorphic. In the category of sets a terminal object is any set
$\{*\}$ with just a single element.  In this case a morphism is just
a map, and hence a morphism $\{*\}\map X$ picks out a unique element
of $X$.} in $\SetB$ defined by $1(W):=\{*\}$ (the set with one
element) for all $W\in\B$.  Specifically, for each $\a\in\UP$ we
define $\tilde\chi^{A^d}(\a):1\map\Om$ by specifying its components
$\tilde\chi^{A^d}(\a)_{W_0}:1(W_0)\map\Om(W_0)$ to be
\beq
	\tilde\chi^{A^d}(\a)_{W_0}(*):=\chi^{A^d}_{W_0}(\a)
									\label{tilde-chiAda=}
\eeq
where the right hand side is given by \eq{Def:chiAd}. In turn, this
produces a map\footnote{We are exploiting here the fact that the constant
presheaf functor $\Delta:\Set\map\SetB$ is left adjoint
to the `global sections' functor
$\G:\SetB\map\Set$ where, for any object $F$ in $\SetB$, we have $\G
F:=\Hom_\SetB(1,F)$. This adjointness relation gives rise to a
natural isomorphism $\Hom_\SetB(\Delta S,F)
\simeq \Hom_\Set(S,\G F)$ for any set $S$. In our case the set $S$ is
$\UP$ and the functor $F$ is $\Om$; thus the isomorphism of interest is
$\Hom_\SetB(\Delta\UP,\Om)\simeq\Hom_\Set(\UP,\G\Om)$. The element in
$\Hom_\SetB(\Delta\UP,\Om)$ with which we are concerned is $\chi^{A^d}$
and its image in $\Hom_\Set(\UP,\G\Om)$ is what we have denoted
$\tilde\chi^{A^d}$. Thus $\tilde\chi^{A^d}(\a)\in\G\Om=\Hom_\SetB(1,\Om)$.}
\beq
		\tilde\chi^{A^d}:\UP\map\Hom_\SetB(1,\Omega)
									\label{tilde-chiAd}
\eeq 
where $\Hom_\SetB(1,\Omega)$ denotes the set of morphisms from $1$
to $\Om$ in the topos category $\SetB$.

	By these means, to each global proposition $\gp{\a}{A^d}$
we can associate a corresponding `valuation' morphism
\beq
		V\gp{\a}{A^d}:1\map\Om		\label{VaAd}
\eeq
where $V\gp{\a}{A^d}:=\tilde\chi^{A^d}(\a)$ is a global element of
$\Om$; \ie $V$ is a map from global propositions to global elements.
In normal set theory, a map from $\{*\}$ (the terminal object in the
category of sets) to a set $X$ picks out a unique element of $X$,
and \eq{VaAd} can be regarded as the analogue of this procedure in
the category $\SetB$ of varying sets.  Thus \eq{VaAd} encapsulates
the idea that in our topos interpretation of the
consistent-histories formalism, a `generalised truth value' is
associated to each global proposition $\gp{\a}{A^d}$---namely the
global element $V\gp{\a}{A^d}:1\map\Om$ of $\Om$.

	Referring to \eq{VaAd} as a `valuation' seems to imply that it
preserves some logical structure on the propositions $\gp{\a}{A^d}$.
However, we do not have any such structure {\em a priori\/}:
rather, the intention of \eq{VaAd} is to use the Heyting algebra
structure of $\Om$ to {\em define\/} a logical algebra on the global
propositions $\gp{\a}{A^d}$---a goal that can be achieved by
associating each such second-level proposition with the
corresponding global element of $\Om$. For example, if
$V\gp{\a}{A^d}:1\map\Om$ and $V\gp{\b}{A^d}:1\map\Om$ are global
elements of $\Om$ corresponding to the global propositions
$\gp{\a}{A^d}$ and $\gp{\b}{A^d}$ respectively then the global
proposition ``$\gp{\a}{A^d}$ {\em and\/} $\gp{\b}{A^d}$'' is
associated\footnote{Note that ``$\gp{\a}{A^d}$ {\em and\/}
$\gp{\b}{A^d}$'' is not itself of the form $\gp{\g}{A^d}$ for any
$\g\in\UP$. It is thus more accurate to think of the propositions
$\gp{\a}{A^d}$, $\a\in\UP$ as {\em generators\/} of a logical
algebra.} with the global element of $\Om$ defined by the chain
\beq
1\stackrel{<V\gp{\a}{A^d},V\gp{\b}{A^d}>}{\longrightarrow}\Om\times\Om
\stackrel{\land}{\longrightarrow}\Om
\eeq 
where $\land:\Om\times\Om\map\Om$ is the `and' operation in the
Heyting algebra structure on $\Om$. This is a rather sophisticated
analogue of the treatment of $\gp{\a}{p}\land\gp{\b}{q}$ by the
expression \eq{Def:Vmand} in the context of our earlier discussion
of second-level propositions in a classical theory.

	Note however that the map $\tilde\chi^{A^d}$ in
\eq{tilde-chiAd} is not one-to-one, and hence neither is the
valuation map $\gp{\a}{A^d}\mapsto V\gp{\a}{A^d}$ that associates a
global element of $\Om$ with each $\gp{\a}{A^d}$.  Thus---analogous
again to our discussion in section \ref{Subsec:second-level} of
classical second-level propositions---we are lead to define two
global propositions as being {\em $d$-semantically equivalent\/} if
they are associated with the same global element of $\Om$ (with a
given decoherence function $d$): properly speaking, it is only to
the equivalence classes of such propositions that the logical
algebra applies.

	For example, although in the construction \eq{Def:chiAd} the
quantity $\Delta\UP$ is regarded purely as a {\em set\/} and the
orthoalgebra structure plays no {\em a priori\/} role,
nevertheless---since $W$ is a Boolean subalgebra of $\UP$---if
$\a\in W$ then $\neg\a\in W$ and {\em vice versa\/}. Hence
$\a\not\in W$ if and only if $\neg\a\not\in W$, which implies that,
for all $\a\in\UP$,
\beq
		\chi^{A^d}_{W_0}(\a)=\chi^{A^d}_{W_0}(\neg\a).
						\label{chiA(a)=chiA(nega)}
\eeq
Thus $\gp{\a}{A^d}$ and $\gp{\neg\a}{A^d}$ are $d$-semantically
equivalent\footnote{I am grateful to Pen Maddy for the gnomic remark
that this conclusion is consistent with proposition 4.0621 in
Wittgenstein's {\em Tractatus\/}.} for any $\a\in\UP$
and for all decoherence functions $d$.

\subsection{The semantic values for $\gp{\a}{p}$}
Finally we are in a position to treat the main goal of the
paper, namely propositions of the type $\gp{\a}{p}$---``the
probability of history proposition $\a$ being true is $p$''.  All we
have to do is to supplement the requirement of $d$-accessibility
with the additional condition $d(\a,\a)=p$.  Thus I propose to
interpret the global proposition $\gp{\a}{p}$ by specifying it to
have the following `localised' form:

\begin{defn}
\item[] The proposition $\gp{\a}p$ is $d$-{\em accessible from a
window $W$\/} if (i) $\a$ is $d$-accessible from $W$; and (ii)
$d(\a,\a)=p$.
\end{defn}

	Let $A^{d,p}(W)$ denote the set of all propositions $\a\in\UP$
such that $\gp{\a}p$ is $d$-accessible from $W$:
\beq
	A^{d,p}(W):=\{\,\a\mid\exists\, W'\supseteq W\mbox{ s.t. }
			 W'\in\Bd,\a\in W',\aand d(\a,\a)=p\,\}.
\eeq
These sets obey the basic condition $W_1\leq W_2$ implies
$A^{d,p}(W_1)\subseteq A^{d,p}(W_2)$ and hence, for each $p\in[0,1]$,
$A^{d,p}:=\{A^{d,p}(W),W\in\B\}$ is a varying set over $\B$. 

	The varying set $A^{d,p}$ is a subobject of the constant varying
set $\Delta\UP$, and hence for each decoherence function $d$ (an
analogue of the state $\s$ that arises in \eq{Def:chimp}) and each
$p\in[0,1]$ we get the crucial characteristic morphism
$\chi^{d,p}:\Delta\UP\map\Om$ in $\SetB$ whose components are the maps
$\chi^{d,p}_{W_0}:\Delta\UP(W_0)\map\Om(W_0)$, $W_0\in\B$, defined
by
\beqa
	\chi^{d,p}_{W_0}(\a):=\left\{\ba{ll}
						\{\,W\subseteq W_0\mid\exists\, W'\supseteq W
							\mbox{ with }W'\in\Bd\aand \a\in W'\,\}
								&\mbox{if $d(\a,\a)=p$;}\\[4pt]
						\emptyset& \mbox{otherwise.}
							\ea
						\right.					\label{Def:chidp}
\eeqa
The right hand side of \eq{Def:chidp} is a genuine
sieve and is to be regarded as the semantic value at stage $W_0$ of
the global proposition $\gp{\a}{p}$ ``the state of affairs
represented by the history proposition $\a\in\UP$ has probability
$p$ of occurring''.

	As was the case with $\gp{\a}{A^d}$, the global proposition
$\gp{\a}p$ can be associated with a global element
$\tilde\chi^{d,p}(\a):1\map\Om$ whose components
$\tilde\chi^{d,p}(\a)_{W_0}:1(W_0)\map\Om(W_0)$ are defined as
$\tilde\chi^{d,p}(\a)_{W_0}(*):=\chi^{d,p}_{W_0}(\a)$. Putting together
these various results we finally arrive at the desired `valuation
morphism'
\beq
	V^d\gp{\a}{p}:1\map\Om			\label{Vdap}
\eeq
whose components $V^d\gp{\a}{p}_{W_0}:1(W_0)\map\Om(W_0)$ are given
by
\beq
	V^d\gp{\a}{p}_{W_0}(*)=\left\{\ba{ll}
						\{\,W\subseteq W_0\mid\exists\, W'\supseteq W
							\mbox{ s.t. }W'\in\Bd\aand \a\in W'\,\}
								&\mbox{if $d(\a,\a)=p$;}\\[4pt]
						\emptyset& \mbox{otherwise.}\label{VdapW0}
							\ea
						\right.		
\eeq

	The topos result \eq{VdapW0} should be compared with the
simple expression \eq{Def:Vm} that applies in a more conventional
probability theory. Once again we see the `fuzzy-set' nature of
the topos scheme in the sense that at any particular stage
$W_0$ the proposition $\gp{\a}{p}$ may be assigned a semantic value
other than $0:=\emptyset$ or $1:=\up{W_0}$.

	As in the earlier example of the second-level propositions
$\gp{\a}{A^d}$, it is appropriate to define two global propositions
$\gp{\a}{p}$ and $\gp{\b}{q}$ to be `$d$-semantically-equivalent' if
they are associated with the same global element of $\Om$, \ie if
their semantic values are equal in all windows. For example, it is
clear from \eq{VdapW0} that, for all decoherence functions $d$, the
second-level propositions $\gp{\a}{p}$ and $\gp{\neg\a}{1-p}$ are
$d$-semantically equivalent for all $\a\in\UP$ and all $p\in[0,1]$.
This is because if $\a$ belongs to some window $W'$ then so does
$\neg\a$. Furthermore, if $W'$ is $d$-consistent then
$d(\a,\neg\a)=0$ and hence, by additivity of the decoherence
function $d$,
\beq
	1=d(1,1)=d(\a\oplus\neg\a,\a\oplus\neg\a)=d(\a,\a)+d(\neg\a,\neg\a)  
\eeq
which shows that $d(\neg\a,\neg\a)=1-d(\a,\a)$.

	By these means, the (equivalence classes of) global
propositions of the type $\gp{\a}{p}$ generate a logical algebra
as a subalgebra of the Heyting algebra on the set
$\Hom_\SetB(1,\Om)$ of global elements of $\Om$ in $\SetB$. The
expression \eq{Vdap}---with its component version
\eq{VdapW0}---represents the final form of our analysis of the
logical structure of the consistent-history propositions ``the
probability that $\a\in\UP$ is realised is $p$'' in the context of
topos theory.

\section{Conclusions}
A key ingredient in the consistent-histories formulation of quantum
theory is the existence of $d$-consistent sets of propositions. We
have argued that, in the approach where a preferred set is {\em
not\/} selected once and for all, the ensuing many world-views
semantics can be described mathematically with a topos-theoretic
framework based on the idea of sets varying over the
partially-ordered set $\B$ of all Boolean subalgebras of the
orthoalgebra of history propositions. In particular, we have seen
how a global proposition such as ``the probability of the history
proposition $\a$ being true is $p$'' can be interpreted in a way
that identifies any window $W_0\in\B$ as a potential `stage' and
where the semantic values at each such stage lie in a Heyting
algebra.  This situation stems from the central claim that each
global proposition can be identified with a global element
$1\map\Om$ of the space $\Om$ of truth values in the topos category
$\SetB$.  Propositions of this type in the consistent-histories
programme include $\gp{\a}{A^d}$, $\gp{\a}p$ and (as discussed in
the second appendix) $\gp{\a}{U^d}$.  The collection of $d$-semantic
equivalence classes of all\footnote{Mixed propositions---for
example, ``$\gp{\a}{U^d}$ {\em and } $\gp{\b}p$''---are allowed.}
such propositions generates a logical structure that is inherited
from that of $\Om$. It should be emphasised once more that, in
practice, the space $\UP$ may simply be the set of projection
operators on some Hilbert space, in which case the analysis of the
crucial poset $\B$ is a viable concrete task.

	The general conclusion of this paper is that topos methods
provide a natural mathematical framework in which to discuss the
inner logical structure that lies behind ideas of many windows, or
world-views, in the quantum theory of histories. One general aim of
this approach is to avoid the instrumentalism that dominates much
conventional thinking about quantum theory although---as is not
infrequently the case---it is difficult to give a simple physical
picture of what the formalism means in these circumstances. However,
if we accept the idea that `classical realism' is associated in some
way with a Boolean algebra of propositions, then we have to say
that---because of its intrinsic Heyting algebra logic---the `many
windows' interpretation of the decoherent histories formalism
corresponds to a type of neo-realism that, on the one hand, is more
complicated and subtle than the simple realism of classical physics
but which, on the other hand, does not go as far as the
non-distributive structure that characterises quantum logic proper.
Of course, this does not affect the the fact that the underlying
orthoalgebra $\UP$ of history propositions {\em is\/} a genuine
quantum logic.

	In the context of `many world-views' it is worth noting that
the concept of a proposition being `$d$-accessible' from a window
$W$ clearly extends to Boolean subalgebras in general: \ie we can
say that a window $W'$ is $d$-accessible from a window $W$ if there
exists a $d$-consistent window $W''$ that contains both $W'$ and
$W$; a single proposition $\a$ is then $d$-accessible from $W$ in
the sense of \eq{Def:AdW} if and only if the window
$\{0,1,\a,\neg\a\}$ is $d$-accessible from $W$ in the sense just
described.  However, this is just the consistent-histories analogue
of the idea of `relative possibility' introduced by Kripke (1963) in
his original study of the semantics of intuitionistic modal logic
(see also Loux, 1979).  This suggests that modal concepts such as
`necessity' or `possibility' should find a natural home in the
quantum formalism of consistent histories, but this remains a topic
for future work.

	The topos-theoretic ideas used in the present paper are rather
elementary, being essentially restricted to the theory of presheafs
on a poset, and there is a lot more to the subject than that.
However, even at the simple level of the theory of varying sets it
seems clear that the ideas discussed here could find applications in
other areas of quantum theory where some type of contextuality
arises. An example might be the idea of `relational quantum theory'
that has been actively developed recently by several authors; for
example Crane (1995), Smolin (1995) and Rovelli (1995). It also
seems possible that the well-known contextuality of truth values in
standard quantum theory (\ie the Kochen-Specker theorem) could be
explored profitably from this perspective.

\bigskip
\noindent
{\bf Acknowledgements}

\noindent
I am most grateful to Jeremy Butterfield, Jonathan Halliwell, Martin
Hyland, Noah Linden and Stephen Schreckenberg for stimulating
discussions and/or helpful comments on earlier drafts of this paper. I
am also grateful to Steve Vickers for introducing me to the
literature on topos theory.

\bigskip
\appendix
\section{An Alternative Approach}
I shall sketch here a method whereby it {\em is\/} possible to
assign the set ({\em c.f.} \eq{VpB0-1}) 
\beq
	\G_\a^d(W_0):=\{\,W\subseteq W_0\mid W\in\Bd \aand \a\in W\,\}
											\label{Def:Gad}
\eeq
as the semantic value at stage $W_0$ of the global proposition
$\gp{\a}{p}$ (on the supposition that $d(\a,\a)=p$) even though the
right hand side of \eq{Def:Gad} is not a sieve on $W_0$ in
$(\B,\leq)$.

	The first observation is that---assuming for simplicity that
$d(\a,\a)=p$ and $d(\b,\b)=q$---if $\G^d_\a(W_0)$ and $\G^d_\b(W_0)$
{\em were\/} to be the semantic values of propositions $\gp{\a}{p}$
and $\gp{\b}{q}$ respectively at stage $W_0$ then the global
proposition ``$\gp{\a}{p}$ {\em or\/}$\gp{\b}{q}$'' would presumably
be represented at stage $W_0$ by the set
\beqa
	\G^d_\a(W_0)\cup \G^d_\b(W_0)&=&
		\{\,W\subseteq W_0\mid W\in\Bd, \a\in W\,\}\cup
			\{\,W\subseteq W_0\mid W\in\Bd, \b\in W\,\}\nonumber\\
	&=&\{\,W\subseteq W_0\mid W\in\Bd 
				\aand(\a\in W \oor \b\in W)\,\}	\nonumber\\
	&=&\{\,W\subseteq W_0\mid W\in\Bd\aand\{\a,\b\}\cap W\neq\emptyset\,\}.
									\label{Ga-and-Gb}
\eeqa

	Since the right hand side of \eq{Ga-and-Gb} is not itself of the form
$\G^d_\g(W_0)$ we do not have algebraic closure. However, the
structure of \eq{Ga-and-Gb} is suggestive and leads to the idea of
defining the `trapped' sets
\beq
		T^d_F(W_0):=\{\,W\subseteq W_0\mid W\in\Bd\aand F\cap
					W\neq\emptyset\,\}				\label{Def:TF}				
\eeq
where $F$ is any finite set of propositions from $\UP$. Note that
$\G^d_\a(W_0)\equiv T^d_{\{\a\}}(W_0)$. 

	The collection of all such sets is closed under the union
operation since
\beq
		T^d_F(W_0)\cup T^d_G(W_0)=T^d_{F\cup G}(W_0)
\eeq
although under intersections we have
\beq
	T^d_F(W_0)\cap T^d_G(W_0)=\{\,W\subseteq W_0\mid W\in\Bd\aand (F\cap
		W\neq\emptyset\;\&\; G\cap W\neq\emptyset)\,\}.
\eeq
Closure can be re-established by defining the collections of
subalgebras
\beqa
\lefteqn{T^d_{F_1,F_2,\ldots,F_m}(W_0) := }		\label{TFm}\\
		& & \{\,W\subseteq W_0\mid W\in B_d\aand
				(F_1\cap W\neq\emptyset\;\&\; F_2\cap W\neq\emptyset,
			\ldots,\;\&\; F_m\cap W\neq\emptyset)\,\}\nonumber
\eeqa
where $F_1,F_2,\ldots,F_m$ is any finite collection of finite sets
of propositions in $\UP$. 

	A simple way of using these sets to generate a logical algebra
follows from the following observation. There is a well known
topology (the Vietoris topology) that can be placed on the set of
all closed subsets of a topological space and which involves
trapping sets, rather as in \eq{Def:TF}. Motivated by what is done
in the Vietoris situation the natural procedure in our case is to
define a topology $\tau_d$ on $\B$ by taking as a subbasis the
collection of all sets of the form $T^d_F(W)$ as $F$ ranges of all
finite subsets of $\UP$ and $W$ ranges over all Boolean
subalgebras of $\UP$. The open sets of the topological space
$(\B,\tau_d)$ can then serve as the semantic values of our system.

	Notice that this procedure does indeed produce a logical
structure since the collection of open sets in any topological space
is always a Heyting algebra. However, this is rather far from our
original idea of presheafs on the partially-ordered set $(\B,\leq)$
and needs to be treated as a separate theory.

\section{Unrealisable Propositions}
The `localised' form of a new, and rather simple, global proposition
``$\a$ is $d$-unrealisable'' (denoted $\gp{\a}{U^d}$) is given by
the following definition.

\begin{defn}
\item[] A proposition $\a\in\UP$ is $d$-{\em unrealisable in
a window $W$\/} if (i) $W$ is $d$-consistent; and (ii) $\a\not\in W$.
\end{defn}
Thus we can define
\beq
		U^d(W):=\{\,\a\mid W\in\Bd\aand \a\not\in W\,\} =
		\left\{\ba{ll}
					\UP-W&\mbox{if $W\in\Bd$}\\
					 \emptyset&\mbox{if $W\not\in\Bd$}
				\ea
		\right.
\eeq
as the set of all propositions\footnote{Note that, according to this
definition, the $0$ and $1$ history propositions are never
$d$-unrealisable.} that are $d$-unrealisable in the window $W$.
We note that:
\be
\item If $W_1\leq W_2$ then $U^d(W_1)\subseteq U^d(W_2)$ and
hence, unlike the case for $R^d$, the collection
$U^d:=\{U^d(W),W\in\B\}$ is a genuine varying set over the
poset $(\B,\leq)$ if the maps $U^d_{W_1W_2}:U^d(W_1)\map U^d(W_2)$,
$W_1\leq W_2$, are defined to be the subset inclusions.

\item The object $U^d$ in $\SetB$ can be regarded as a
subobject of the constant varying set $\Delta\UP$ in $\SetB$. The
associated characteristic morphism is
\beqa
	\chi^{U^d}_{W_0}:\Delta\UP(W_0)&\map& \Om(W_0)			\nonumber\\
		\a\quad&\mapsto& \{\,W\geq W_0\mid \a\in U^d(W)\,\}\nonumber\\
				&\ =& \{\,W\subseteq W_0\mid W\in\Bd\aand \a\not\in W\,\}.
					\label{chiUd}
\eeqa
\ee

	As was the case with $\gp{\a}{A^d}$, the new global proposition
$\gp{\a}{U^d}$ can be associated with a global element of the Heyting
algebra $\Om$ via the morphism
\beq
		\tilde\chi^{U^d}(\a):1\map\Om
\eeq	
whose components are defined by
$\tilde\chi^{U^d}(\a)_{W_0}(*):=\chi^{U^d}_{W_0}(\a)$ ({\em c.f.}
\eq{tilde-chiAda=}).  There is an associated valuation morphism
$V\gp{\a}{U^d}:1\map\Om$ where
$V\gp{\a}{U^d}:=\tilde\chi^{U^d}(\a)$. We also note that,
analogously to \eq{chiA(a)=chiA(nega)},
\beq
	\chi^{U^d}_{W_0}(\a)=\chi^{U^d}_{W_0}(\neg\a)
\eeq
for all windows $W_0$ and all $\a\in\UP$. Thus $\gp{\a}{U^d}$ and
$\gp{\neg\a}{U^d}$ are $d$-semantically equivalent for any
$\a\in\UP$ and for all decoherence functions $d$.

	Finally we remark that one might have tried to use these
results to resolve the `realisability' problem by defining a
proposition $\a\in\UP$ to be $d$-realisable in a window if it is
not $d$-unrealisable there. This involves taking the
negation\footnote{Simply taking the set-theoretic complement of
each $U^d(W)$, $W\in\B$, will not produce a proper element of
$\SetB$ since the resulting sets $\{\a\mid\a\not\in U^d(W)\}$
have the wrong behaviour with respect to $W_1\subseteq W_2$. This
was mentioned earlier in the context of \eq{Def:negA(p)}.} of
the variable set $U^d$ in the appropriate Heyting algebra of
subobjects of the constant variable set $\Delta\UP$.  The result
is the variable set $\neg U^d:=\{\neg U^d(W),W\in\B\}$ where
\beqa
	\neg U^d(W)&=&\{\,\a\mid \forall\, 
				W'\geq W, \a\not\in U^d(W')\,\}		\nonumber\\
			&=&\{\,\a\mid\forall\,W'\subseteq W, 
				W'\not\in\Bd\oor\a\in W'\,\}.		\label{negId}
\eeqa
However, since $W':=\{0,1\}$ is a $d$-consistent coarse-graining of any
window $W$ it follows that $\neg U^d(W)=\{0,1\}$---a rather
small number of `realisable' propositions!

\bigskip\bigskip\noindent
{\bf REFERENCES}

\begin{trivlist}

\item[] Bell, J.L. (1988).
  {\em Toposes and Local Set Theories: {A}n Introduction}.
  Clarendon Press, Oxford.

\item[] Butterfield, J. (1996). Whither the minds?
  {\em Brit. J. Phil. Science.}, {\bf 47} 200--221.

\item[] Crane, L. (1995).
  Clocks and categories: {I}s quantum gravity algebraic?
  {\em J. Math. Phys.}, {\bf 36 } 6180--6193.

\item[] Dowker, H.F. and Kent, A. (1995).
  Properties of consistent histories.
  {\em Phys. Rev. Lett.}, {\bf 75} 3038--3041.

\item[] Dowker, H.F. and Kent, A. (1996).
  On the consistent histories approach to quantum mechanics.
  {\em J. Stat. Phys}, {\bf 82} 1575--1646.

\item[] Dummett, M. (1959).
  Truth. {\em Proc. Aristotelian Soc.}, {\bf 59} 141--162.

\item[] Foulis, D.J., Greechie, R.J. and R\"uttimann, G.T. (1992).
  Filters and supports in orthoalgebras.
  {\em Int. J. Theor. Phys.}, {\bf 31} 789--807.

\item[] Gell-Mann, M. and Hartle, J. (1990a).
  Alternative decohering histories in quantum mechanics.
  In {\em Proceedings of the 25th International Conference
  on High Energy Physics, Singapore, August, 1990}, K.K.~Phua and
  Y.~Yamaguchi, eds. World Scientific, Singapore.

\item[] Gell-Mann, M. and Hartle, J. (1990b).
  Quantum mechanics in the light of quantum cosmology.
  In  {\em Complexity, Entropy and the Physics of
  Information, SFI Studies in the Science of Complexity, {Vol.
  VIII}}, W.~Zurek, ed. Addison-Wesley, Reading.

\item[] Griffiths, R.B. (1984).
  Consistent histories and the interpretation of quantum mechanics.
  {\em J. Stat. Phys.}, {\bf 36} 219--272.

\item[] Griffiths, R.B. (1993).
  {\em Found. Phys.}, {\bf 23} 1601.

\item[] Griffiths, R.B. (1996).
  Consistent histories and quantum reasoning. quant-ph/9606004.

\item[] Hardegree, G.M. and Frazer, P.J. (1982).
  Charting the labyrinth of quantum logics: {A} progress report.
  In {\em Current Issues in Quantum Logic}, E.G.~Beltrametti
  and B.V.~van Frassen, eds. Plenum Press, London.

\item[] Halliwell, J. (1995).
	A review of the decoherent histories approach to quantum
	mechanics. In {\em Fundamental Problems in Quantum Theory},
	D.~Greenberger, ed. 

\item[] Hartle, J. (1991). The quantum mechanics of cosmology.
  In {\em Quantum Cosmology and Baby Universes}, S.~Coleman, J.~Hartle,
  T.~Piran, and S.~Weinberg, eds. World Scientific, Singapore.

\item[] Hartle, J. (1995).
  Spacetime quantum mechanics and the quantum mechanics of spacetime.
  In  {\em Proceedings of the 1992 Les Houches School,
  Gravitation and Quantisation}, B.~Julia and J.~Zinn-Justin, eds.
  Elsevier Science, Amsterdam.

\item[] Isham, C.J. (1994).
  Quantum logic and the histories approach to quantum theory.
  {\em J. Math. Phys.}, {\bf 35} 2157--2185.

\item[] Isham, C.J. (1995). 
  Quantum logic and decohering histories. 
  In {\em Topics in Quantum Field Theory}, D.~H.~Tchrakian ed. 
  World Scientific, Singapore.

\item[] Isham, C.J. and Linden, N. (1994).
  Quantum temporal logic and decoherence functionals in the histories
  approach to generalised quantum theory.
  {\em J. Math. Phys.}, {\bf 35} 5452--5476.

\item[] Isham, C.J. and Linden, N. (1995).
  Continuous histories and the history group in generalized quantum
  theory. {\em J. Math. Phys.}, {\bf 36} 5392--5408.

\item[] Isham, C.J. and Linden, N. (1996).
	Information-entropy and the space of decoherence functions in
	generalised quantum theory.
	In preparation. 

\item[] Isham, C.J., Linden, N. and Schreckenberg, S. (1994).
  The classification of decoherence functionals: an analogue of
  {G}leason's theorem. {\em J. Math. Phys.}, {\bf 35} 6360--6370.

\item[] Kent, A. (1996). Consistent sets contradict. gr-qc/9604012.

\item[] Kripke, S. (1963). Semantical considerations on modal logic.
  {\em Acta Philosophica Fennica}, {\bf 16\/} 83--94.

\item[] Lawvere, F.W. (1975)
  Continuously variable sets: algebraic geometry = geometric logic.
  In {\em Proceedings Logic Colloquium Bristol 1973}.
  North-Holland, Amsterdam.

\item[] Loux, M.J. (1979). {\em The Possible and the Actual}.
  Cornell University Press, London.

\item[] Mac{L}ane, S. and Moerdijk, I. (1992).
  {\em Sheaves in Geometry and Logic: {A} First Introduction to Topos
  Theory}. Springer-Verlag, London.

\item[] Omn\`es, R. (1988a).
  Logical reformulation of quantum mechanics. {I.} {F}oundations.
  {\em J. Stat. Phys.}, {\bf 53} 893--932.

\item[] Omn\`es, R. (1988b)
  Logical reformulation of quantum mechanics. {II.} {I}nterferences and
  the {E}instein-{P}odolsky-{R}osen experiment.
  {\em J. Stat. Phys.}, {\bf 53} 933--955.

\item[] Omn\`es, R. (1988c).
  Logical reformulation of quantum mechanics. {III.} {C}lassical limit
  and irreversibility.  {\em J. Stat. Phys.}, {\bf 53} 957--975.

\item[] Omn\`es, R. (1989).
  Logical reformulation of quantum mechanics. {IV.} {P}rojectors in
  semiclassical physics.  {\em J. Stat. Phys.}, {\bf 57} 357--382.

\item[] Omn\`es, R. (1990)
  From {H}ilbert space to common sense: {A} synthesis of recent
  progress in the interpretation of quantum mechanics.
  {\em Ann. Phys. (NY)}, {\bf 201} 354--447.

\item[] Omn\`es, R. (1992)
  Consistent interpretations of quantum mechanics.
  {\em Rev. Mod. Phys.}, {\bf 64} 339--382.

\item[] Rovelli, C. (1996). Relational quantum theory.
  {\em Int. J. Theor. Phys.}. In press.

\item[] Smolin, L. (1995).
  The {B}ekenstein bound, topological quantum field theory, and
  pluralistic quantum cosmology.  gr-qc/9508064.

\item[] Wittgenstein, L. (1966). {\em Tractatus Logico-Philosophicus}.
  Routledge {\&} Kegan Paul, London.
\end{trivlist}

\end{document}